\documentclass[journal=jctc,manuscript=letter]{achemso}

\usepackage{amsmath,amssymb}
\usepackage{bm}
\usepackage{braket}
\usepackage{upgreek}

\newcommand{\hh}{\mathcal{H}}

\title{Neural Quantum States Based on Selected Configurations}

\author{Marco Julian Solanki}
\affiliation{ETH Z\"urich, Department of Chemistry and Applied Biosciences, Vladimir-Prelog-Weg 2, CH-8093 Z\"urich, Switzerland}

\author{Lexin Ding}
\affiliation{ETH Z\"urich, Department of Chemistry and Applied Biosciences, Vladimir-Prelog-Weg 2, CH-8093 Z\"urich, Switzerland}

\author{Markus Reiher}
\affiliation{ETH Z\"urich, Department of Chemistry and Applied Biosciences, Vladimir-Prelog-Weg 2, CH-8093 Z\"urich, Switzerland}
\email{mreiher@ethz.ch}

\begin{document}


\begin{abstract}
Neural quantum states (NQS) provide a flexible and highly expressive parameterization of wave functions for strongly correlated problems in quantum chemistry. Despite rapid advances in network architectures, the evaluation of electronic energies remains almost exclusively based on variational Monte Carlo (VMC). While VMC is effective for structured systems such as spin chains, its accuracy and efficiency for electronic Hamiltonians are hindered by sharply peaked distributions, stochastic gradient noise, and slow convergence with sample size. 
In this letter, we assess the capability of NQS-VMC to efficiently capture correlation in electronic ground states by comparing it to a recently developed NQS-based selected configuration (NQS-SC) approach. 
We set up a systematic comparison of the ground-state optimizations obtained with NQS-VMC and NQS-SC for molecular systems dominated by either static or dynamical correlation. The comparison demonstrates a clear advantage of NQS-SC over NQS-VMC in both energy accuracy and wave-function coefficients, particularly for statically correlated molecules. Moreover, NQS-SC exhibits robust systematic improvability, whereas NQS-VMC does not. These findings position NQS-SC as the new default approach over NQS-VMC for electronic structure calculations. We further observe that neither NQS-SC nor NQS-VMC can efficiently capture dynamical correlation, highlighting the need for future hybrid methods, such as multiconfigurational perturbation theories built on top of NQS solutions. 
\end{abstract}

\textbf{Introduction.} Neural quantum states (NQS) have been proposed as a novel ansatz for solving the many-body Schr\"odinger equation \cite{2017-CarleoTroyer}. Leveraging the expressive power of artificial neural networks allows NQS to deliver a compact representation for strongly correlated wave functions. Theoretical analyses and numerical studies on lattice Hamiltonians have demonstrated that NQS can represent highly entangled states using a number of network parameters that scales only polynomially with system size \cite{deng2017entanglement, zakari2025comment}. These results on the representational power of NQS have motivated a rapidly growing body of literature exploring increasingly sophisticated network architectures ranging from restricted Boltzmann machines to autoregressive models \cite{glasser2018neural, choo2018rbm, 2019-NBF, hibat2020recurrent, sharir2020deep, kochkov2021graph, schmale2022arcnn, zhang2023transformer, reh2023cnn, denis2025bosoncnn, lange2025gutzwiller, lange2025transformer, ibarra2025auto}, some of which have been applied to \textit{ab initio} quantum chemistry \cite{2020-FermiNet, 2020-PauliNet, 2020-RBM, yang2020RBM, 2022-NAQS, 2023-MADE, pfau2024accurate, 2024-NBF, 2025-BDG-RNN, 2025-RetNet, 2023-QiankunNet, 2025-QiankunNet-Backflow}. However, the practical scope of NQS for molecular systems remains limited to about 50 electrons in 40 spatial orbitals \cite{2023-MADE}, as a central challenge is the efficient access to the exponentially many wave-function coefficients required for accurate energy evaluation and optimization. 

Currently, the predominant method for evaluating the energy of an NQS is variational Monte Carlo (VMC) sampling. For early NQS, the sampling method of choice was Metropolis-Hastings sampling \cite{2017-CarleoTroyer, 2020-RBM}, while many later NQS implementations applied autoregressive sampling \cite{2022-NAQS, 2023-MADE, 2023-QiankunNet, 2025-RetNet, 2025-BDG-RNN}. The motivation behind this shift was the observation that molecular wave functions are typically dominated by a small subset of configurations (with a long tail of configurations with small weights), leading to high rejection rates during Metropolis-Hastings sampling. For example, Ref.~\citenum{2020-RBM} reported acceptance rates as low as 0.1\% for calculations on \(\text{H}_\text{2}\text{O}\) in a 6-31G basis set. The efficiency of Metropolis-Hastings sampling is further limited (i) by the need to prune intermediate samples to obtain approximately independent and identically distributed samples and (ii) by the need for a burn-in period for the sampled Markov chains to equilibrate \cite{MCMC}. Conceptually, these issues can be resolved with autoregressive sampling, whereby the neural network outputs conditional probabilities that, when sequentially sampled, yield samples from the full probability distribution \cite{2022-NAQS}. This results in a rejection-free sampling procedure that produces a sequence of uncorrelated samples. 

Yet, also autoregressive sampling has its shortcomings. First, it requires specific network architectures, which include an amplitude network to model the probability amplitude of each configuration (such as a recurrent neural network or decoder-only transformer) and a separate phase network for predicting the complex phase of the coefficients \cite{2022-NAQS, 2023-MADE, 2023-QiankunNet, 2025-RetNet, 2025-BDG-RNN}. Second, autoregressive models are unable to directly enforce crucial symmetries in electronic systems, such as fixed particle numbers and magnetizations, thereby necessitating the ad hoc masking/discarding of unphysical configurations \cite{2022-NAQS, 2023-MADE, 2023-QiankunNet, 2025-RetNet, 2025-BDG-RNN}. Finally, the ordering of the orbitals will affect the ground-truth conditional probability distributions, which, in turn, may affect the model's ability to learn them. Ref.~\citenum{2023-MADE}, for example, used an entanglement-localizing ordering to improve the performance of their model. However, this means that autoregressive models, unlike fully-connected NQS, cannot claim invariance to orbital orderings as a potential advantage over competing tensor network methods. 

To tackle these issues of VMC for electronic ground states, Li \textit{et al.} proposed an NQS-based selected configuration (NQS-SC) approach for energy evaluations \cite{2023-RBM-SC}. This approach takes strong inspiration from traditional selected configuration interaction (SCI) methods (see, e.g., Refs.~\citenum{bender1969studies, schriber2016communication, HCI, liu2016ici, zimmerman2017incremental, chilkuri2021comparison}), which have previously been applied to challenging molecular systems governed by both strong correlation \cite{ASCI, SCI-Algos, HCI, SHCI} and system size \cite{SCI-Algos}. These traditional SCI methods typically select configurations from outside the selected variational space based on information beyond the ansatz for the truncated wave function, such as perturbation theories or criteria that closely resemble them \cite{CIPSI, ASCI, HCI, SHCI}. NQS-SC, by contrast, selects configurations based on the probability amplitudes predicted by the NQS ansatz, which is defined not only on the selected variational space but also on the external space in which new configurations can arise. The predicted wave-function coefficients of these external configurations enter the SC energy implicitly via the so-called local energy. 

It has been shown that the non-stochastic nature of NQS-SC can lead to more stable optimizations and higher accuracy in the electronic ground-state energy than NQS-VMC \cite{2023-RBM-SC, 2024-NBF, 2025-NBF}. However, as a truncated method, NQS-SC can recover only static correlation. Namely, it can only account for a small fraction of configurations from the exponentially large Hilbert space that make significant contributions to the exact solution. Moreover, the current formulation of the NQS-SC energy as a truncated sum of local energies is not variational, which, if left unchecked, could lead to an underestimation of the energy error \cite{2023-RBM-SC}.

Hence, a rigorous inspection of NQS-SC relative to NQS-VMC is lacking, but needed to unlock the potential of the NQS parametrization. Therefore, we here evaluate (final) NQS-SC energies variationally in two different ways to facilitate a fair comparison with NQS-VMC. As a result, we can clearly demonstrate the general advantage of NQS-SC over NQS-VMC, which supports the role of SC as a new standard component in NQS parametrizations of electronic ground states.

\textbf{Theory.} We start by considering the Hilbert space $\hh$ of $N$ electrons ($N_{\alpha}$ spin-up and $N_{\beta}$ spin-down) in $2L$ spin-orbitals ($L$ spin-up and $L$ spin-down). The second-quantized electronic Hamiltonian can be expressed as
\begin{equation}
\hat{H} = \sum_{i,j=1}^{2L} h_{ij}\hat{a}_{i}^\dag\hat{a}_{j}^{\phantom{\dag}} + \frac{1}{2}\sum_{i,j,k,l=1}^{2L} V_{ijkl}\hat{a}_{i}^\dag\hat{a}_{j}^\dag\hat{a}_{k}^{\phantom{\dag}}\hat{a}_{l}^{\phantom{\dag}},
\end{equation}
where $h_{ij}$ and $V_{ijkl}$ are one- and two-electron integrals (the nucleus-repulsion terms have been omitted for the sake of simplicity) and $\hat{a}^{(\dagger)}_{i}$ are fermionic annihilation (creation) operators that annihilate (create) an electron in the $i$-th spin-orbital. The exact ground state wave function in such a basis is obtained by variationally minimizing the energy \(E^\mathrm{FCI}\) associated with the full configuration interaction (FCI) ansatz
\begin{equation}
    |\Psi^{\rm FCI}\rangle = \sum_{\ket{\bm{n}}\in\mathcal{B}}\Psi(\bm{n})\ket{\bm{n}},
\end{equation}
where \(\ket{\bm{n}} = \ket{n_1, n_2, \dots, n_{2L}}\) are occupation number vectors (ONVs) (with \(n_i = 0,1\) representing the occupation of the $i$-th spin-orbital), which jointly form a complete basis \(\mathcal{B}\) of \(\hh\). 

The total dimension of the Hilbert space $\hh$ is determined combinatorically as \(\textrm{dim}(\hh)=\binom{L}{N_\alpha}\binom{L}{N_\beta}\), which steeply constrains the largest numbers of electrons and orbitals that one can describe with FCI in practice, requiring approximations for all but the smallest of systems. We therefore consider 
an NQS $\Psi_{\bm{\theta}}(\bm{n})$ to denote the neural network approximant to the mapping \(\bm{n}\mapsto\Psi(\bm{n})\) for a neural network with trainable parameters $\bm{\theta}$. 

In the NQS-VMC framework, $n_{\rm sample}$ (non-unique) configurations are sampled to approximate the probability distribution induced by the neural network 
\begin{equation}
    P_{\bm \theta}({\bm n}) = \frac{|\Psi_{\bm\theta}({\bm n})|^2}{\sum_{\ket{\bm m}\in\mathcal{B}}|\Psi_{\bm \theta}({\bm m})|^2}.
\end{equation}
There are many methods to generate samples, with the most widely employed options being (i) the Metropolis-Hastings algorithm, which proposes transitions between configurations and accepts or rejects them based on energetic and probabilistic considerations, and (ii) the autoregressive approach, which decomposes the full probability distribution \(P_{\bm \theta}({\bm n})\) into a sequence of univariate conditional probabilities. The performance of specific samplers can be sensitive to hyperparameter choices, such as the proposal distribution and number of Markov chains for Metropolis-Hastings and the orbital ordering for autoregressive sampling. For a conclusive evaluation of NQS-VMC performance, we therefore need to choose the most accurate, albeit computationally most expensive, exact Monte Carlo (EMC) sampling method. 

In EMC, ONV samples are drawn directly from the distribution $P_{\bm \theta}({\bm n})$, thereby eliminating all sources of error beyond statistical sampling noise. Once the multiset of samples $\mathcal{S}_{\rm MC}$ has been generated, the energy of the NQS is approximated as
\begin{equation}
    E_{\bm{\theta}} = \sum_{\ket{\bm{n}} \in \mathcal{B}} P_{\bm{\theta}}({\bm n}) E_{\bm \theta}^{\rm loc}({\bm n}) \approx \sum_{\ket{\bm n} \in \mathcal{S}_{\rm MC}} E^{\rm loc}_{\bm{\theta}}({\bm n}) \equiv E^{\rm MC}_{\bm \theta}, \label{eqn:en_vmc}
\end{equation}
where $E_{\bm \theta}^{\rm loc}({\bm n})$ is the local energy defined as
\begin{equation}
    E_{\bm \theta}^{\rm loc}({\bm n}) = \sum_{\ket{\bm m} \in \mathcal{B}} \langle {\bm n} | \hat{H} | {\bm m}\rangle \frac{\Psi_{\bm{\theta}}(\bm{m})}{\Psi_{\bm{\theta}}(\bm{n})}.
\end{equation}
In EMC, the most deciding factor for accuracy is the number of samples $n_{\rm sample}$. According to the central limit theorem, 
the standard error $\delta E$ of the energy scales as $1/\sqrt{n_{\rm sample}}$ \cite{MCMC}. Elementary probability theory also dictates that a configuration $\ket{\bm n}$ with probability amplitude $P_{\bm \theta}({\bm n})$ can only be expected to be accessed if $n_{\rm sample}$ is of the order of at least $P_{\bm \theta}({\bm n})^{-1}$. As the probability distributions of electronic ground states are often dominated by only a small number of important configurations (static correlation), they are, by construction, followed by a long tail of configurations with vanishing weights but considerable collective contribution to the ground-state energy (dynamical correlation). VMC calculations, therefore, often require a vast number of samples to achieve an energy accuracy that can support chemical accuracy of about 1 kcal/mol \cite{2022-NAQS, 2023-QiankunNet, 2025-RetNet}. 

By contrast, in the SC framework, the weighted sum in Equation (\ref{eqn:en_vmc}) is truncated to a set of $n_{\rm select}$ configurations $\mathcal{S}_{\rm select}$, which contains the configurations with the highest probability amplitudes predicted by the neural network. After the network parameters are updated, new configurations are selected from the extended set $\mathcal{S}_{\rm extend} = \hat{H}\mathcal{S}_{\rm select}$ consisting of configurations connected to $\mathcal{S}_{\rm select}$ by the electronic Hamiltonian. In practice, we construct $\mathcal{S}_{\rm expand}$ as a random subset of $\mathcal{S}_{\rm extend}$ of size $n_{\rm expand}$ and $n_{\rm select}$ new configurations are selected from the union $\mathcal{S}_{\rm select}\,\cup\,\mathcal{S}_{\rm expand}$ to form the new $\mathcal{S}_{\rm select}$. The size of $\mathcal{S}_{\rm select}$ is thereby maintained between iterations. The initial \(\mathcal{S}_{\rm select}\) consists of the Hartree--Fock configuration as a seed, and all configurations connected to it (or a random subset of them, if the resulting size of the space would otherwise exceed \(n_\mathrm{select}\)). Motivating this choice is the observation that low-excitation-rank CI wave functions can be a good initial guess for many molecular systems if the variationally optimized Hartree--Fock determinant is taken as a reference for the excitations. 

Given a set $\mathcal{S}_{\rm select}$ and some \(\ket{\bm{n}}\in\mathcal{S}_{\rm select}\), \(P_{\bm \theta}({\bm n})\) can be approximated as
\begin{equation}
P^\mathrm{SC}_{\bm{\theta}}(\bm{n}) = \frac{\vert\Psi_{\bm{\theta}}(\bm{n})\vert^2}{\sum_{\ket{\bm{m}}\in\mathcal{S}_{\rm select}} \vert\Psi_{\bm{\theta}}(\bm{m})\vert^2}. \label{eq:sc_prob}
\end{equation}
The electronic energy can then be estimated as
\begin{equation}
E_{\bm \theta} \approx \sum_{\ket{\bm{n}}\in\mathcal{S}_{\rm select}} P^\mathrm{SC}_{\bm{\theta}}(\bm{n})E_{\bm \theta}^{\rm loc}({\bm n}) \equiv E_{\bm \theta}^{\rm SC}. \label{eq:sc_energy}
\end{equation}
As already noted by Ref.~\citenum{2023-RBM-SC}, and in contrast to the EMC energy estimate in Equation (\ref{eqn:en_vmc}), this energy estimate is non-variational. In our calculations, we recover variationality in two ways. First, the symmetric evaluation of the electronic energy is calculated as
\begin{equation}
E_{\bm \theta} \approx \frac{\sum_{\ket{\bm{m}},\ket{\bm{n}}\in\mathcal{S}_{\rm select}} {\Psi_{\bm{\theta}}(\bm{m})}^{\ast}\braket{\bm{m}|\hat{H}|\bm{n}}\Psi_{\bm{\theta}}(\bm{n})}{\sum_{\ket{\bm{n}}\in\mathcal{S}_{\rm select}}\vert\Psi_{\bm{\theta}}(\bm{n})\vert^2} \equiv E_{\bm \theta}^{\textrm{SC-SYM}}. \label{eq:sc_energy_exact}
\end{equation}
This amounts to an energy evaluation with respect to the NQS $|\Psi_{\rm \bm{\theta}}\rangle$ truncated on \(\mathcal{S}_\mathrm{select}\). The variationality of this estimate follows from the variational principle for admissible trial wavefunctions.
Second, the variationally optimal energy on $\mathcal{S}_{\rm select}$ is calculated through exact diagonalization. 
This can be viewed as the result of selected configuration interaction (SCI)
\begin{equation}
    E^{\textrm{SCI}} = \min_{|\Phi\rangle\,\in\,\textrm{span}(\mathcal{S}_{\rm select})} \frac{\langle \Phi |\hat{H}|\Phi\rangle}{\langle \Phi|\Phi\rangle}. \label{eq:sci_energy}
\end{equation}
For the purpose of training, $E_{\bm \theta}^{\rm SC}$ is still the more suitable objective function, because it is computationally inexpensive and the wave-function coefficients of the configurations not only in $\mathcal{S}_{\rm select}$ but also $\mathcal{S}_{\rm extend}$ implicitly enter it through the local energies. The discrepancy between $E_{\bm \theta}^{\rm SC}$ and $E_{\bm \theta}^{\rm SC-SYM}$ should be reduced as more and more configurations are selected and finally agree in the limit of \(n_\mathrm{select} = \mathrm{dim}(\hh)\). 

Another crucial factor that determines the performance of a particular NQS implementation is its network architecture. For this, we chose the neural backflow (NBF) architecture first introduced in Ref.~\citenum{2019-NBF} as a variational approximation to the Fermi--Hubbard model. This architecture was later applied to \textit{ab initio} quantum chemistry in Ref.~\citenum{2024-NBF}. Recently, it has also been considered in combination with a Jastrow factor \cite{2025-NBF} and through the replacement of its feedforward layers with an encoder-only transformer \cite{2025-QiankunNet-Backflow}. It was furthermore shown to be capable of learning wave functions of even highly entangled fermionic systems with volume-law scaling \cite{zakari2025comment}. Beyond fermions, NBF has also been adapted for bosons and applied to anharmonic vibrational structure calculations \cite{ding2025modal}.

We briefly review the NBF model. We start by considering a single Slater determinant (SD) state
\begin{equation}
    |\Psi^{\rm SD}\rangle = \hat{f}^\dagger_1 \hat{f}^\dagger_2 \cdots \hat{f}^\dagger_N |0\rangle,
\end{equation}
where $\hat{f}^\dagger_j = \sum_{i=1}^{2L} C_{ij} \hat{a}^\dagger_i$ creates an electron in the orbital specified by the $j$-th column of the orbital coefficient matrix $\bm{C}$. $|\Psi^{\rm SD}\rangle$ can be expanded in the basis of ONVs of the form $|\bm{n}\rangle \!=\! \prod_{i=1}^{2L} (\hat{a}^\dagger_i)^{n_i}|0\rangle$ as
\begin{equation}
\begin{split}
    |\Psi^{\rm SD}\rangle &= \left(\sum_{i_1}^{2L} C_{i_11} \hat{a}^\dagger_{i_1}\right)\left(\sum_{i_2}^{2L} C_{i_22} \hat{a}^\dagger_{i_2}\right)\cdots \left(\sum_{i_N}^{2L} C_{i_NN} \hat{a}^\dagger_{i_N}\right)|0\rangle
    \\
    &= \sum_{i_1i_2\cdots i_N}^{2L} C_{i_11}C_{i_22}\cdots C_{i_NN} \hat{a}^\dagger_{i_1}\hat{a}^\dagger_{i_2}\cdots \hat{a}^\dagger_{i_N}|0\rangle 
    \\
    &= \sum_{i_1<i_2<\cdots<i_N}^{2L} \left( \sum_{\pi \in S_N} {\rm sgn}(\pi) C_{i_{\pi(1)}1}C_{i_{\pi(2)}2}\cdots C_{i_{\pi(N)}N} \right) \hat{a}^\dagger_{i_1}\hat{a}^\dagger_{i_2}\cdots \hat{a}^\dagger_{i_N}|0\rangle,
\end{split}
\end{equation}
where $\pi$ are elements of the permutation group $S_N$ and $\rm{sgn}(\pi)\!\in\!\{\pm1\}$ is the sign of the permutation. In other words, the expansion coefficient $\Psi^{\rm SD}(\bm{n})$ can be expressed as the determinant
\begin{equation}
    \Psi^{\rm SD}(\bm{n}) = \langle \bm{n}|\Psi^{\rm SD}\rangle = {\rm det} \,\bm{C}({\bm n}), \label{eqn:SD}
\end{equation}
where $\bm{C}({\bm n})$ is a square matrix consisting of rows of the orbital coefficient matrix $\bm{C}$ whose indices $i$ satisfy $n_i\!=\!1$. 

Equation \eqref{eqn:SD} reveals the structure of the expansion coefficients $\Psi^{\rm SD}(\bm{n})$ of the SD state. Namely, they are coordinated through the determinants of different submatrices of the \textit{same} orbital coefficient matrix $\bm{C}$. An NQS directly learning the $\Psi^{\rm SD}(\bm{n})$ faces the nontrivial task of learning this determinant structure, even though $|\Psi^{\rm SD}\rangle$ is a simple mean field state without any electron correlation.

The NBF model makes use of this insight and expands it by learning, for each ONV $\bm{n}$, a \textit{different} orbital coefficient matrix $\bm{C}^{(\bm{n})}$. The coefficient $\Psi^{\rm NBF}(\bm{n})$ is then given by the backflow (i.e., ONV-dependent) determinant
\begin{equation}
    \Psi^{\rm NBF}(\bm{n}) = {\rm det}\,\bm{C}^{(\bm{n})}(\bm{n}),
\end{equation}
where $\bm{C}^{(\bm{n})}(\bm{n})$ is again understood to consist of rows of $\bm{C}^{(\bm{n})}$ whose indices $i$ satisfy $n_i\!=\!1$.
This way, the task of learning the determinant structure is offloaded from the network. At the same time, $\Psi^{\rm NBF}(\bm{n})$ can, thanks to the ONV-dependence of $\bm{C}^{(\bm{n})}$ and in contrast to $\Psi^{\rm SD}(\bm{n})$ in Equation \eqref{eqn:SD}, vary freely and therefore introduce electron correlation.

In practice, NBF can be set to learn more than one orbital coefficient matrix for each ONV. An ONV $\bm{n}$ is passed through \(K\geq 1\) feedforward layers with widths \(d_k\), each performing a linear transformation \(\bm{y}^{(k)} \!=\! \bm{W}^{(k)}\bm{x}^{(k-1)} + \bm{b}^{(k)}\) and a non-linear activation \(\bm{x}^{(k)} \!=\! \bm{\sigma}(\bm{y}^{(k)})\). 
The weights \(\bm{W}^{(k)}\!\in\!\mathbb{R}^{d_{k}\times d_{k-1}}\) and biases \(\bm{b}^{(k)}\!\in\!\mathbb{R}^{d_{k}}\) are network parameters to be optimized. 
The final (activation-less) layer's output \(\bm{y}^{(K)}\) has length \(d_K = D\times2L\times N\) and is reshaped into \(D\) matrices 
with dimensions \(2L\times N\) denoted as \(\bm{C}^{(\bm{n})}_1, \bm{C}^{(\bm{n})}_2, \dots, \bm{C}^{(\bm{n})}_D \in \mathbb{R}^{2L\times N}\). Using these $D$ matrices, the coefficient $\Psi^{\rm NBF}_{\bm{\theta}}(\bm{n})$ predicted by the NBF network for the ONV $\bm{n}$ is given by
\begin{equation}
\Psi^{\rm NBF}_{\bm{\theta}}(\bm{n}) = \sum_{i=1}^D \det {\bm{C}^{(\bm{n})}_i}(\bm{n}),
\end{equation}
where $\bm{\theta}$ collects the weights $\bm{W}^{(k)}$ and biases $\bm{b}^{(k)}$.

In the following, the NBF model trained with EMC sampling and selected configurations will be referred to as NBF-VMC and NBF-SC, respectively. All energies are measured in Hartree (Ha).

\begin{figure}
\centering
\includegraphics[width=\textwidth]{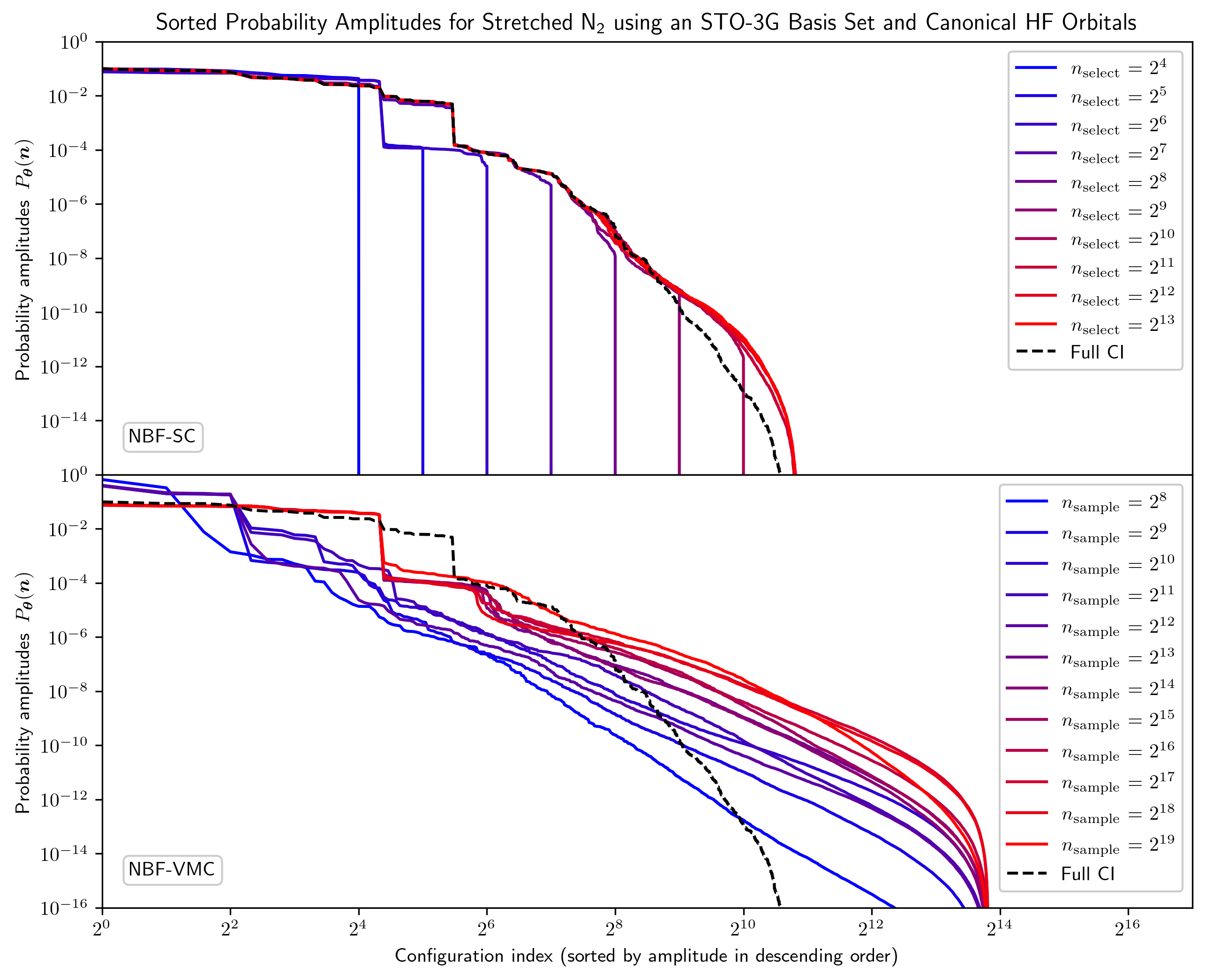}
\caption{Sorted probability amplitudes from the NBF model trained with varying \(n_\mathrm{select}\) and \(n_\mathrm{sample}\) using selected configurations and exact Monte Carlo sampling compared to the FCI solution for stretched \(\text{N}_\text{2}\) using an STO-3G basis set and canonical HF orbitals (14,400 total configurations).}
\label{fig:C-N2s-STO-3G-HF}
\end{figure}

\begin{figure}
\centering
\includegraphics[width=\textwidth]{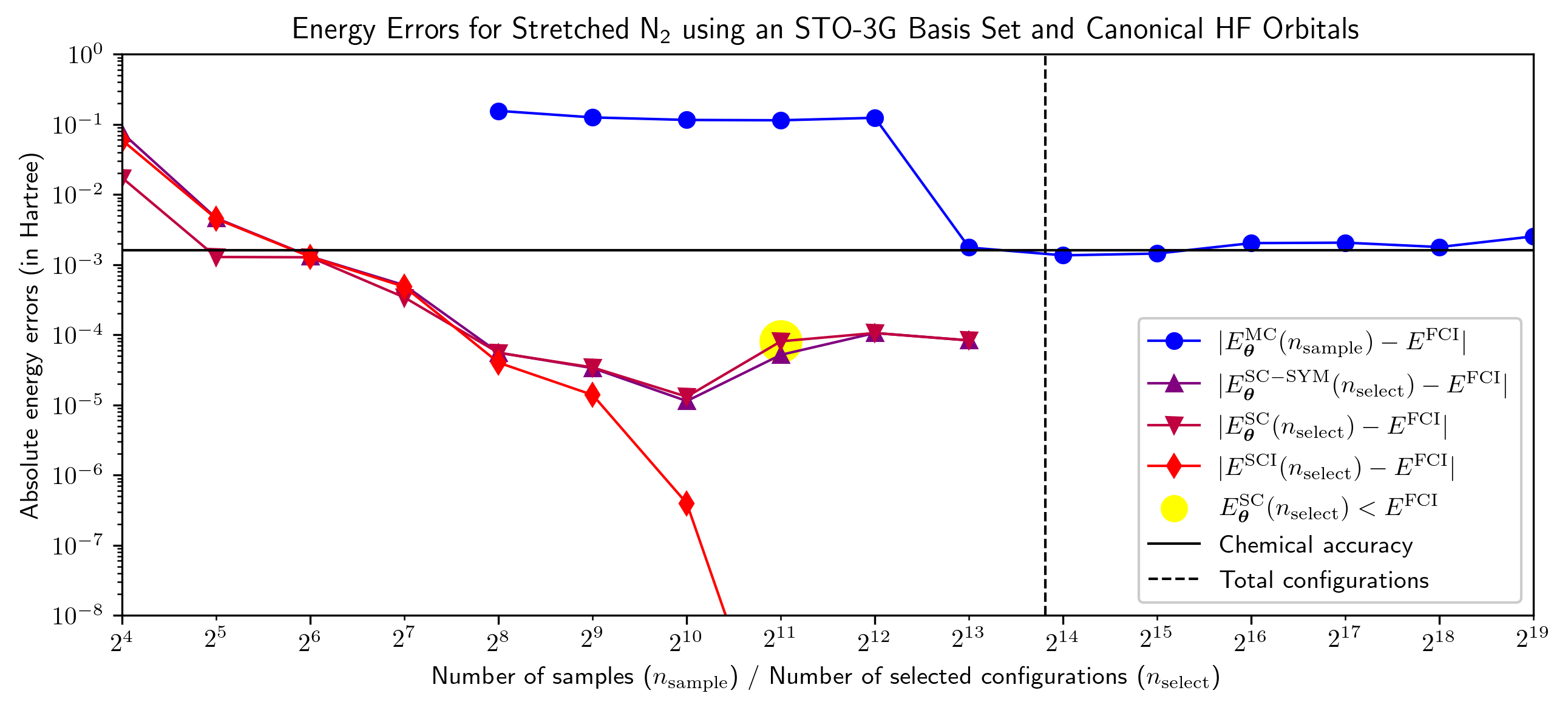}
\caption{Energy errors for the NBF solutions plotted in Figure \ref{fig:C-N2s-STO-3G-HF}. For \(n_\mathrm{select} = 2^{11}\), \(\vert E^\mathrm{SCI}(n_\mathrm{select}) - E^\mathrm{FCI}\vert\) drops down to approx.\ \(5\cdot 10^{-11}\) Ha where it remains for subsequent \(n_\mathrm{select}\).}
\label{fig:E-N2s-STO-3G-HF}
\end{figure}

\textbf{Stretched $\mathbf{N_2}$.} We start our comparison with a prototypical example with strong static correlation, namely the stretched \(\text{N}_\text{2}\) molecule (with a bond length of \(2.25\) \r{A}).
To limit our memory consumption during training and retain our ability to compute an FCI reference solution, we chose a minimal STO-3G basis set. We present the probability amplitudes predicted and energy errors incurred by NBF-VMC and NBF-SC in Figures \ref{fig:C-N2s-STO-3G-HF} and \ref{fig:E-N2s-STO-3G-HF}, respectively.

We see that even with a small number of selected configurations, NBF-SC can effortlessly identify the \(20\) most dominant (largest amplitude) configurations. It struggles slightly more with the amplitude plateau induced by the \(24\) next most dominant configurations, but manages to identify them once \(n_\mathrm{select}\geq 2^7\).

While NBF-SC can predict the highest-contributing amplitudes reasonably well, it struggles with describing the tail of the distribution (where \(P_{\bm{\theta}}(\bm{n})\lesssim 10^{-8}\)). In terms of energies, \(E^\mathrm{SC-SYM}_{\bm{\theta}}\) drops below chemical accuracy (i.e., below an energy error of about \(1.6\) mHa)
for \(n_\mathrm{select} = 2^6\) and improves until reaching its minimum error at \(11.4\) \(\upmu\)Ha for \(n_\mathrm{select} = 2^{10}\). It is particularly notable that while \(E^\mathrm{SCI}\) remains close to \(E^\mathrm{SC-SYM}_{\bm{\theta}}\) for \(n_\mathrm{select}\leq 2^{8}\), \(E^\mathrm{SCI}\) drops significantly below \(E^\mathrm{SC-SYM}_{\bm{\theta}}\) for \(n_\mathrm{select}\geq 2^{9}\) reaching an error of approximately \(5\cdot 10^{-11}\) Ha \(\approx 0\) for \(n_\mathrm{select} \geq 2^{11}\). The remarkable accuracy of  \(E^\mathrm{SCI}\) signals that NQS-SC successfully found the correct configurations, although the predicted coefficients are not highly accurate, as \(E^\mathrm{SC-SYM}_{\bm{\theta}}\) actually increases for \(n_\mathrm{select} > 2^{10}\). It is also noteworthy that for \(n_\mathrm{select} = 2^{11}\), \(E^\mathrm{SC}_{\bm{\theta}}\) falls below \(E^\mathrm{FCI}\), indicating that the non-variationality of this energy estimate can pose an issue.

As for NBF-VMC, the \(20\) most dominant configuration amplitudes are only recovered once \(n_\mathrm{sample}\geq 2^{13}\) are considered. This sudden improvement in the predicted probability amplitudes as \(n_\mathrm{sample}\) is increased from  \(2^{12}\) to \(2^{13}\) corresponds to a significant drop in the energy error from \(124\) mHa to \(1.76\) mHa. Chemical accuracy is finally reached at \(n_\mathrm{sample} = 2^{14} =\) 16,384, which is on the same order as the dimension of the Hilbert space \(\mathrm{dim}(\hh) =\) 14,400. From this point on, the NBF-VMC energy error does not significantly improve any further. This behaviour may be attributed to the fact that NBF-VMC never recovers the large amplitudes for the \(21\mathrm{st}\) to \(44\mathrm{th}\) most dominant configurations in the FCI solution. At the same time, NQS-VMC completely fails to describe the tail of the amplitude distribution even as $n_{\rm sample}$ is increased.

\begin{figure}
\centering
\includegraphics[width=\textwidth]{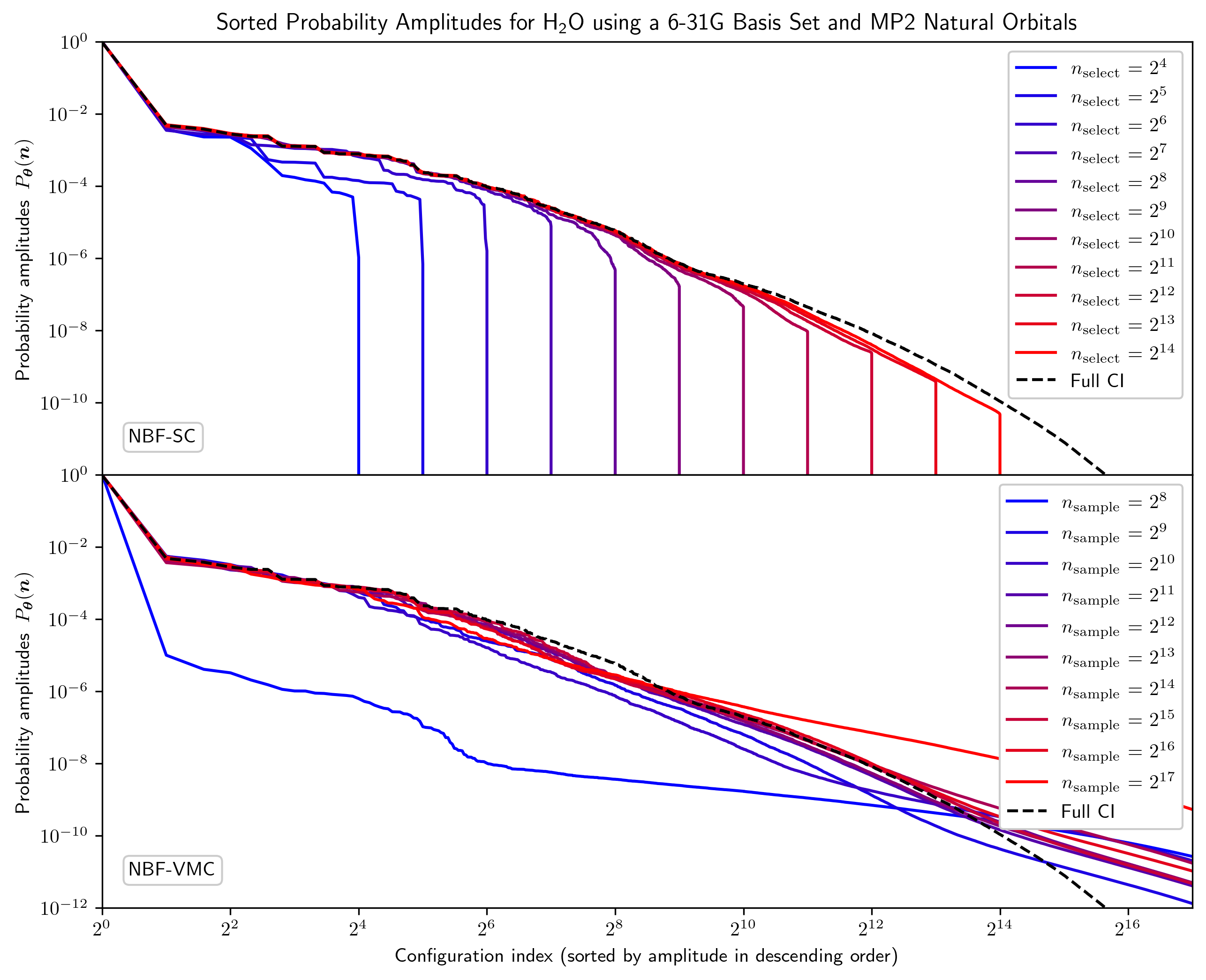}
\caption{Sorted probability amplitudes from the NBF model trained with varying \(n_\mathrm{select}\) and \(n_\mathrm{sample}\) using selected configurations and exact Monte Carlo sampling compared to the FCI solution for \(\text{H}_\text{2}\text{O}\) using a 6-31G basis set and MP2 natural orbitals (1,656,369 total configurations).}
\label{fig:C-H2O-6-31G-NO}
\end{figure}
\begin{figure}
\centering
\includegraphics[width=\textwidth]{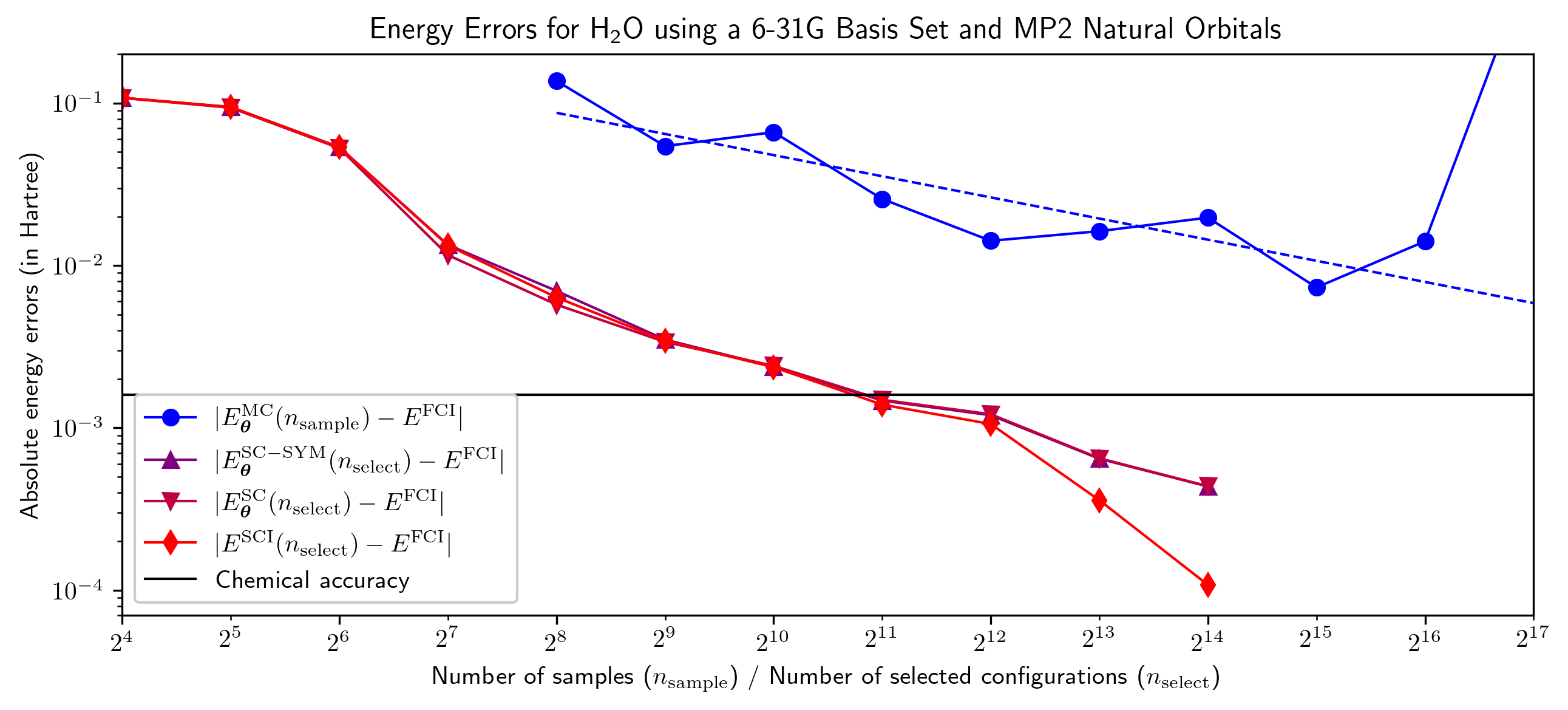}
\caption{Energy errors for the NBF solutions plotted in Figure \ref{fig:C-H2O-6-31G-NO}. \(\vert E_{\bm{\theta}}^\mathrm{MC}(n_\mathrm{sample}) - E^\mathrm{FCI}\vert\) for \(n_\mathrm{sample} = 2^{17}\) is approx.\ \(837\) mHa. The blue dashed trendline was calculated by linear least-squares in log-log space, ignoring the outlier value of \(\vert E_{\bm{\theta}}^\mathrm{MC}(n_\mathrm{sample}) - E^\mathrm{FCI}\vert\) for \(n_\mathrm{sample} = 2^{17}\).
}
\label{fig:E-H2O-6-31G-NO}
\end{figure}

\textbf{$\mathbf{H_2O}$.} The \(\text{H}_\text{2}\text{O}\) molecule in a 6-31G basis set is taken as an example of a system with comparatively strong dynamical correlation. Comparisons of the probability amplitudes predicted by NBF-VMC and NBF-SC, as well as of the corresponding energy estimates, are provided in Figures \ref{fig:C-H2O-6-31G-NO} and \ref{fig:E-H2O-6-31G-NO}, respectively.

In terms of coefficients, the probability amplitudes predicted by NQS-SC for a given \(n_\mathrm{select}\) tend to start deviating earlier (at lower configuration indices) from the FCI solution for the $\rm H_2O$ molecule compared to the stretched $\rm N_2$ molecule. These deviations can, however, for the most part be counteracted by simply increasing $n_{\rm select}$. In terms of energy, \(E^\mathrm{SC-SYM}_{\bm{\theta}}\) also systematically improves as $n_{\rm select}$ is increased, and finally drops below chemical accuracy for \(n_\mathrm{select} = 2^{11}\). Furthermore, in contrast to stretched \(\mathrm{N}_\mathrm{2}\), \(E^\mathrm{SC}_{\bm{\theta}}\) remains
above \(E^\mathrm{FCI}\) for all calculations. Also, \(E^\mathrm{SC}_{\bm{\theta}}\) and \(E^\mathrm{SC-SYM}_{\bm{\theta}}\) remain close for all $n_{\rm select}$, which means the asymmetric evaluation of the SC energy is not an issue here. However, the energy optimization still struggles for large \(n_\mathrm{select}\). This is indicated by the observation that \(E^\mathrm{SCI}\) visibly drops below \(E^\mathrm{SC-SYM}_{\bm{\theta}}\) for large enough \(n_\mathrm{select}\), meaning the wave-function coefficients predicted by NQS-SC are not optimal for that selected subspace. Overall, Figures \ref{fig:C-H2O-6-31G-NO} and \ref{fig:E-H2O-6-31G-NO} show that the NQS-SC procedure performs reasonably well for systems dominated by dynamical correlation, with systematic improvements in the energy errors, but also at the cost of an increasingly large number of selected configurations, as one would expect.

NBF-VMC struggles even more with $\rm H_2O$ compared to stretched $\rm N_2$. Although the energy error now steadily improves with more samples, it never reaches chemical accuracy. For \(n_\mathrm{sample} = 2^{17}\), the VMC calculations do not even converge, which is not uncommon and presumably due to suboptimal sample batches pushing the model off its convergence trajectory. Changing the sampler seed restores the expected convergence behavior. With a linear regression in log-log space while ignoring the outlier for \(n_\mathrm{sample} = 2^{17}\), we extrapolate \(E^\mathrm{MC}_{\bm{\theta}}\) to fall below chemical accuracy for \(n_\mathrm{sample} \approx\) 2,643,513 \(\approx 2^\mathrm{21.3}\), that is, for NBF-VMC to reach chemical accuracy, it would again take \(n_\mathrm{sample}\) on the order of the dimension of the entire Hilbert space \(\mathrm{dim}(\hh) =\) 1,656,369. This is contrary to the conceptual advantage that NQS-VMC should have over NQS-SC, namely that, as an approach operating over the entirety of \(\hh\), NQS-VMC should be able to better capture dynamical correlation than NQS-SC. In practice, the minimum error obtained with NBF-SC in the presence of dynamical correlation remains over an order of magnitude lower than that achieved by NBF-VMC.

\begin{table}
\centerline{
\begin{tabular}{c|c|c|c|c}
\hline\hline
Molecule & Basis Set & dim$(\hh)$ & \(\hspace{15pt}n_\mathrm{select}\hspace{15pt}\) & \(n_\mathrm{select}\,/\,\text{dim}(\hh)\) \\
\hline
\(\text{N}_\text{2}\) & STO-3G & 14,400 & \(2^7\) & \(0.889\%\) \\
Stretched \(\text{N}_\text{2}\) & STO-3G & 14,400 & \(2^6\) & \(0.444\%\) \\
\(\text{Li}\text{Cl}\) & STO-3G & 1,002,001 & \(2^7\) & \(0.013\%\) \\
\(\text{H}_\text{2}\text{O}\) & 6-31G & 1,656,369 & \(2^{11}\) & \(0.124\%\) \\
\(\text{C}_\text{2}\text{H}_\text{4}\) & STO-3G & 9,018,009 & \(2^{13}\) & \(0.091\%\) \\
\(\text{Li}_\text{2}\text{O}\) & STO-3G & 41,409,225 & \(2^{10}\) & \(0.002\%\) \\
\(\text{H}_\text{2}\text{O}\) & 6-311G & 135,210,384 & \(2^{14}\)\makebox[0pt][l]{ (?)} & \(0.012\%\)\makebox[0pt][l]{ (?)}  \\
\hline\hline
\end{tabular}}
\caption{Overview of the minimal (power-of-two) \(n_\mathrm{select}\) values needed for NBF-SC to drop below chemical accuracy (\(\vert E^\mathrm{SC-SYM}_{\bm{\theta}}(n_\mathrm{select}) - E^\mathrm{FCI}\vert < 1.6\) mHa) for various combinations of molecules and basis sets. \(\mathrm{dim}(\hh)\) provides the total dimensionality of the underlying Hilbert spaces, while \(n_\mathrm{select}\,/\,\mathrm{dim}(\hh)\) denotes the percentage of total configurations that had to be selected to reach chemical accuracy. The result for the \(\text{H}_\text{2}\text{O}\) molecule in a 6-311G basis set is extrapolated, as \(2^{13}\) was the greatest \(n_\mathrm{select}\) accommodated by NetKet before running out of hardware memory resources.}
\label{tab:sc_performance}
\end{table}

In addition to the two example systems discussed above, we extend our analysis of NQS-VMC and NQS-SC to $\rm N_2$ at equilibrium bond length (\(1.0976\) \r{A}), $\rm LiCl$, $\rm C_2H_4$, and $\rm Li_2O$, using experimental geometries from Ref.~\citenum{CCCBDB} and a combination of STO-3G, 6-31G, and 6-311G basis sets (see Table \ref{tab:sc_performance}).
Probability amplitude profiles and energies similar to those presented in Figures \ref{fig:C-N2s-STO-3G-HF}--\ref{fig:E-H2O-6-31G-NO} are provided for all these systems in the Supporting Information. For these additional systems, NBF-SC again shows a consistent systematic improvement with increasing $n_{\rm select}$. By contrast, the optimized energies of NBF-VMC experienced either sudden and non-systematic jumps with increasing $n_{\rm sample}$ or a too slow convergence (often accompanied by sporadic non-convergence) to chemical accuracy. In addition, we observe that \(E^\mathrm{SC}_{\bm{\theta}}\) only drops below \(E^\mathrm{FCI}\) in the case of one additional system besides stretched \(\text{N}_\text{2}\), namely for \(\text{Li}\text{Cl}\) for \(n_\mathrm{select} = 2^7\). This indicates that the non-variationality of \(E^\mathrm{SC}_{\bm{\theta}}\) only manifests itself rarely.

In Table \ref{tab:sc_performance} we present an overview of the minimal \(n_\mathrm{select}\) required for NBF-SC to reach chemical accuracy for all molecules, except for $\rm H_2O$ using a 6-311G basis set, where the minimal required $n_{\rm select}$ was obtained via extrapolation.
Overall, NQS-SC consistently required fewer than \(1\%\) of all configurations to be selected to reach chemical accuracy. Remarkably, the minimal \(n_\mathrm{select}\) for stretched \(\mathrm{N}_\mathrm{2}\) is half of that for \(\mathrm{N}_\mathrm{2}\) at its equilibrium bond length, despite the fact that the former is more strongly correlated. This confirms that NQS-SC is better suited for systems dominated by static rather than dynamical correlation.
Furthermore, enlarging the basis set for \(\rm H_2O\) from 6-31G to 6-311G reduces the ratio $n_{\rm select}\,/\,{\rm dim}(\hh)$ by a factor of \(10\). While this is generally reassuring for the scalability of NQS-SC with basis set size, the rate of decay of this ratio is still too slow so that, with a 6-311G basis set, chemical accuracy could not be reached before hardware memory constraints were hit. However, we note that this limitation is rather a consequence of the current energy evaluation in NetKet, although the approach also suffers from the general drawbacks of selected CI methods demanding more configurations in such cases. 
This further underscores NQS-SC's susceptibility to systems with slowly decaying amplitude distributions.

\textbf{Conclusions.} To summarize, we compared the performance of two NQS frameworks, NQS-VMC and NQS-SC, in the context of \textit{ab initio} quantum chemistry. We found that, despite the dominance of NQS-VMC in the literature, NQS-SC is a more reliable method in terms of both accuracy and systematic improvability. Our detailed analysis of wave-function coefficient profiles obtained with both frameworks demonstrates that NQS-SC can efficiently capture strong static correlation in electronic ground states and, to some extent, dynamical correlation.

By contrast, NQS-VMC generally struggles to incorporate either type of correlation. Both calculated and extrapolated data indicate that NQS-VMC tends to require an enormous number of samples, rivalling or even exceeding the Hilbert space's total dimension to reach chemical accuracy. However, a large selected configuration space is also needed for NQS-SC to describe molecules dominated by dynamical correlation, although the scaling appears to be far less severe. In such regimes, NQS-SC is unlikely to be a competitive method, as dynamical correlation without its static counterpart can be reasonably accounted for by single-reference methods, such as those based on coupled cluster theory \cite{bartlett2007cc,musial2023advanced}.

Our results highlight a key aspect of NQS development: how information, such as energy and gradients, is extracted from the neural network. The default VMC framework fundamentally limits the efficiency and applicability of NQS to large-scale electronic structure calculations, especially in the regime of interest where both strong static and dynamical correlation are crucial for accurate solutions. Such a limitation is independent of the choice of the Monte Carlo sampling method, as we have conducted our comparison using an exact sampling scheme. As such, we anticipate that NQS-SC will emerge as the new default approach for future NQS implementations for electronic structure calculations. Our findings also underscore the importance of developing improved techniques for extracting physical information from NQS, an aspect that has received limited attention to date.

With more advanced neural network architectures for predicting electronic wave functions now available, it is worthwhile to take a step back and re-evaluate the orthogonal aspect of information extraction to maximize the benefits of these advancements. Another potential route to scale up NQS is to confine them to an active space. This approach is already common for strongly but primarily statically correlated ansätze such as matrix product states \cite{baiardi2020dmrg,ma2022dmrg}. 
For these, the missing dynamical correlation is usually recovered through the subsequent application of a perturbation (see, e.g., Refs.~\citenum{finley1998multi, kurashige2011dmrgcaspt2, freitag2017cddmrgnevpt2, battaglia2020extended}) or coupled cluster theory \cite{veis2016tcc, kinoshita2005tcc, morchen2020tcc, feldmann2024casicc}. 
In this context, NQS-SC naturally emerges as a suitable foundation for future hybrid static–dynamical methods.

\textbf{Methods.} Neural networks were trained using NetKet \cite{NetKet, NetKet3}, JAX \cite{JAX}, and Flax \cite{Flax}. Hartree--Fock, MP2, and FCI calculations were performed with PySCF \cite{Libcint, PySCF, PySCF-Recent}. Variational energy evaluations according to Equations (\ref{eq:sc_energy_exact}) and (\ref{eq:sci_energy}) were performed with PyCI \cite{PyCI}. The molecular structures employed (apart from the stretched \(\text{N}_\text{2}\) molecule) are based on published experimental measurements and were retrieved from Ref.~\citenum{CCCBDB}.

\begin{acknowledgement}
LD gratefully acknowledges financial support of ETH Zurich by an ETH Postdoctoral Fellowship.
\end{acknowledgement}


\providecommand{\latin}[1]{#1}
\makeatletter
\providecommand{\doi}
  {\begingroup\let\do\@makeother\dospecials
  \catcode`\{=1 \catcode`\}=2 \doi@aux}
\providecommand{\doi@aux}[1]{\endgroup\texttt{#1}}
\makeatother
\providecommand*\mcitethebibliography{\thebibliography}
\csname @ifundefined\endcsname{endmcitethebibliography}
  {\let\endmcitethebibliography\endthebibliography}{}

\end{document}